\documentclass[aps,prd,10pt,twocolumn,superscriptaddress,showpacs,floatfix]{revtex4-1}
\usepackage{textcomp,gensymb}
\usepackage{hyperref}
\usepackage{graphicx}
\usepackage{amsfonts,amsmath,amssymb,bm,bbm}
\usepackage{color,soul}
\usepackage{slashed}
\usepackage[toc,page]{appendix}
\usepackage{flushend}
\usepackage{physics}
\usepackage{graphicx}
\usepackage{dcolumn}
\usepackage{float}
\usepackage{fontawesome} 
\newcommand{\githubmaster}{\href{https://github.com/ccapp413/DMpropPublic}{\faGithub}}
\graphicspath{{./figures/}}

\hypersetup{
    pdfnewwindow=true,      
    colorlinks=true,       
    linkcolor=blue,          
    citecolor=blue,        
    filecolor=blue,      
    urlcolor=blue        
}

\begin{document}

\title{An Analytic Approach to Light Dark Matter Propagation}

\author{Christopher V. Cappiello}
\email{cvc1@queensu.ca}
\affiliation{Department of Physics, Engineering Physics, and Astronomy, Queen's University, Kingston, Ontario, K7N 3N6, Canada}
\affiliation{Arthur B. McDonald Canadian Astroparticle Physics Research Institute, Kingston ON K7L 3N6, Canada}
\affiliation{Perimeter Institute for Theoretical Physics, Waterloo, Ontario, N2L 2Y5, Canada}

\begin{abstract}
If dark matter interacts too strongly with nuclei, it could be slowed to undetectable speeds in Earth's crust or atmosphere before reaching a detector. For sub-GeV dark matter, approximations appropriate for heavier dark matter fail, necessitating the use of computationally expensive simulations. We present a new, analytic approximation for modeling attenuation of light dark matter in the Earth. We show that our approach agrees well with Monte Carlo results, and can be much faster at large cross sections. We use this method reanalyze constraints on sub-dominant dark matter. \githubmaster
\end{abstract}

\maketitle


\section{Introduction}
Despite making up most of the mass in the universe, the particle nature of dark matter (DM) remains unknown \cite{Bertone:2004pz}. The most frequently studied candidate is the GeV-scale Weakly Interacting Massive Particle (WIMP), but the lack of a discovery after decades of searching has led to increased interest in other candidates~\cite{Jungman:1995df,Duffy:2009ig,Carr:2020xqk,Feng:2010gw,Leane:2018kjk,Marsh:2018zyw,Essig:2022dfa}. These include sub-GeV DM, which could have evaded detection by carrying too little energy to trigger detectors~\cite{Boehm:2003ha,Boehm:2003hm,Fayet:2004bw,Essig:2011nj,Essig:2012yx,Graham:2012su,Knapen:2016cue,An:2017ojc,CRESST:2017ues,Emken:2017hnp,Knapen:2017xzo,Bringmann:2018cvk,Cappiello:2018hsu,Ema:2018bih,LUX:2018akb,Yin:2018yjn,Cappiello:2019qsw,CDEX:2019hzn,CRESST:2019jnq,DAMIC:2019dcn,EDELWEISS:2019vjv,Krnjaic:2019dzc,XENON:2019zpr,DES:2020fxi,Jaeckel:2020oet,Kim:2020bwm,Kim:2020ipj,SENSEI:2020dpa,Smirnov:2020zwf,An:2021qdl,COSINE-100:2021poy,Elor:2021swj,Emken:2021lgc,Jho:2021rmn,Kamada:2021muh,PROSPECT:2021awi,EDELWEISS:2022ktt,Essig:2022dfa,SuperCDMS:2022kgp}. DM with masses 1 MeV $\lesssim M_{\chi} \lesssim$ 1 GeV can be produced in a variety of frameworks, including hidden sector models~\cite{Feng:2008mu,Feng:2008ya,Hooper:2008im,Boddy:2014yra,Foot:2014uba,Kamada:2021muh},
models involving a dark photon~\cite{Holdom:1985ag,Chu:2011be,Davoudiasl:2013jma,Kaplinghat:2013yxa,Foot:2014uba,Dutra:2018gmv}, and strongly interacting particles that undergo 3 to 2 annihilation~\cite{Hochberg:2014dra,Hochberg:2014kqa,Smirnov:2020zwf}. 

Another topic that has attracted recent interest is sub-dominant DM, i.e. particles that make up only a small fraction of DM, which could have evaded detection because of their low flux~\cite{Bottino:1999ei,Duda:2001ae,Boehm:2003ha,Palazzo:2007gz,Zurek:2008qg,Profumo:2009tb,Belanger:2011ww,Boddy:2014yra,Foot:2014uba,Baum:2016oow,DeLuca:2018mzn,Gross:2018zha,Lehnert:2019tuw,Pospelov:2019vuf,Kamada:2021muh,Billard:2022cqd,Li:2022idr,McKeen:2022poo,Prabhu:2022dtm}. The models of Refs.~\cite{Boehm:2003ha,Boddy:2014yra,Foot:2014uba,Kamada:2021muh} fall under both categories, as they can contain multiple DM states, including a subdominant state with mass around 100 MeV. Dark photon models can also produce small relic densities and large cross sections, resulting in a population of subdominant, strongly interacting DM~\cite{McKeen:2022poo}. 

Experimental sensitivity to both light and subdominant DM is limited to relatively large cross sections: for light DM, because the detectors sensitive to it are optimized for low energy threshold rather than large exposure, and for sub-dominant DM, because of its small flux. This makes attenuation of DM in the Earth's crust or atmosphere a serious concern in both cases. For heavy DM, modeling attenuation by assuming straight particle trajectories is a valid and common simplification~\cite{Starkman:1990nj,Kavanagh:2017cru,Bramante:2018tos,EDELWEISS:2019vjv,Foot:2014osa,Bramante:2018qbc,Bramante:2019yss,Bhoonah:2020fys,DEAPCollaboration:2021raj}. But for light DM, although the straight-line assumption may give reasonably accurate results in some cases~\cite{Davis:2017noy,Emken:2018run,Kavanagh:2017cru,Laletin:2019qca}, this is not always true~\cite{Emken:2019tni}, and accurately modeling DM trajectories has previously required detailed Monte Carlo methods~\cite{Emken:2017qmp,Mahdawi:2017utm,Emken:2018run,Cappiello:2019qsw,Chen:2021ifo,DEAPCollaboration:2021raj,Xia:2021vbz,DaMaSCUS,DaMaSCUS-CRUST}. Accurately assessing direct detection limits is crucial when proposing new DM searches, but can be computationally expensive when using Monte Carlo simulations. 

In this Letter, we describe a novel method of modeling DM attenuation, which has applications to direct detection as well as DM capture in planets, the Sun, neutron stars, white dwarfs, and various other media~\cite{Kouvaris:2007ay,Mack:2007xj,Sandin:2008db,Cumberbatch:2010hh,Frandsen:2010yj,Kouvaris:2010jy,Taoso:2010tg,McDermott:2011jp,Zentner:2011wx,Kouvaris:2012dz,Bell:2013xk,Bramante:2013hn,Catena:2015uha,Baryakhtar:2017dbj,Busoni:2017mhe,Acevedo:2019agu,Bramante:2019fhi,Acevedo:2020gro,Garani:2020wge,Leane:2020wob,Banks:2021sba,Kozar:2021iur,Bramante:2022pmn,Coffey:2022eav,Leane:2022hkk,Nguyen:2022zwb}. We start from the more accurate approximation that light DM particles do not follow straight trajectories, but scatter isotropically in the lab frame when colliding with nuclei. Our method computes probabilities of reaching detector depth analytically, without needing to simulate large numbers of particles, making it feasible in the limit of many scatterings. We validate this method through comparison with Monte Carlo results, and compute limits on sub-dominant DM, showing that direct detection experiments exclude significantly less parameter space than claimed in previous work.

\section{Modeling of Attenuation}\label{code description}

\subsection{Monte Carlo Approach}

Monte Carlo codes model DM propagation by simulating the trajectories of individual particles. After a particle is initialized with speed and arrival direction drawn from the DM velocity distribution, its mean free path is computed, and its position incremented by a distance drawn from the corresponding path length distribution. When the particle scatters with a nucleus, the isotope and scattering angle are drawn from the relevant probability distributions, and the velocity is updated. This process is repeated until the particle either reaches detector depth (such codes often neglect backscattering from below the detector), scatters out of the atmosphere, or loses too much energy to be detected. This process is repeated for thousands or millions of particles, until enough statistics are collected to accurately represent the velocity distribution at the detector.

Sophisticated Monte Carlo codes such as DaMaSCUS~\cite{Emken:2017qmp,DaMaSCUS} model the three-dimensional geometry of the Earth and take the DM wind into account, though we neglect both of these effects in our calculations. Because the rotation of the Earth produces a wide temporal variance in arrival direction~\cite{Kavanagh:2017cru}, and light particles scatter through large angles, we expect this effect in the limit of many scatterings to be negligible.

The Monte Carlo approach accurately models the relevant physical processes on a particle-by-particle basis, accounting for fluctuations in scattering angle and number of collisions. The major weakness is that as the number of scatterings increases, the computational power needed to compute the DM event rate in a detector increases rapidly: not only must each particle be propagated through more collisions, but more particles must be propagated in order to obtain sufficient statistics. This causes the computation time to increase exponentially with cross section. This problem can be severe enough to necessitate the use of specialized rare-event techniques, such as importance sampling or importance splitting~\cite{Mahdawi:2017utm,Emken:2019tni}.

\subsection{Convolutional Approach}

We present an alternative to the Monte Carlo approach. Rather than simulating individual particles, we analytically compute the probability of reaching detector depth $z$ in $n$ scatterings. Our approach can be compared to Ref.~\cite{Hooper:2018bfw}, which conservatively modeled attenuation by considering only particles that do not scatter, and Ref.~\cite{Kavanagh:2016pyr}, which considered particles that scatter at most once. In contrast, this work computes the velocity distribution for all particles, up to an arbitrary number of collisions.

We start by asking what fraction of DM particles reach a depth $z$ without scattering. For a mean free path $l$, the probability density for traveling a distance $x$ before scattering is

\begin{equation}
    P_{\text{initial}}(x) = \frac{1}{l}e^{-x/l}\,.
\end{equation}
Assuming that DM arrives isotropically (from above), and integrating over the zenith angle, the probability of reaching depth $z$ without scattering is

\begin{equation}
    P_0(z) = \frac{1}{l}\Gamma(0,z/l) = -\frac{1}{l}Ei(-z/l)\,,
\end{equation}
where $\Gamma(i,x)$ is the upper incomplete $\gamma$ function, and $Ei(x)$ is the exponential integral.

Next, we need the probability density of reaching depth $z$ after scattering exactly once. This is the probability of reaching an intermediate depth $z'$, times the probability of going from $z'$ to $z$ without scattering, integrated over all values of $z'$. In other words, a convolution of two probability distributions:

\begin{equation}
    P_1(z) = \int P_0(z')P(z-z')dz'\,.
\end{equation}
We worked out $P_0(z)$ above, so we just need the second piece. As mentioned in the Introduction, we now make the simplifying assumption that DM scatters isotropically in the lab frame when colliding with nuclei. The relationship between the lab-frame and CM-frame (center of momentum) scattering angles is given by $\tan \theta_{lab} = \frac{\sin \theta_{CM}}{\cos \theta_{CM} + M_{\chi}/M_A}$, so as long as $M_{\chi} \ll M_A$, isotropic scattering in the CM frame leads to approximately isotropic scattering in the lab frame. By analogy with the case for an isotropic flux, the resulting form for $P(z-z')$ is

\begin{equation}
    P(z-z') = \frac{1}{2l}\Gamma(0,|(z-z')/l|)\,,
\end{equation}
where the factor of 2 and the absolute value account for the fact that the DM could scatter in either the upward or downward direction. The resulting probability to reach detector depth in one scattering is
\begin{equation}
    P_1(z) = \int_0^{\infty} \frac{1}{l}\Gamma(0,z')\frac{1}{2l}\Gamma(0,|z-z'|/l)dz'\,.
\end{equation}

This process can now be iterated: to compute the probability of reaching detector depth after $n$ scatterings, we simply compute

\begin{equation}
    P_n(z) = \int_0^{\infty} P_{n-1}(z')\frac{1}{2l}\Gamma(0,|z-z'|/l)dz'\,.
\end{equation}
The Gamma function and probability distribution both fall off rapidly as $z' \rightarrow \infty$, so the upper integration limit can be set to a finite value that is large compared to the detector depth with minimal loss of accuracy. Setting the upper limit to exactly the detector depth is equivalent to not considering backscattering from below the detector. Throughout this Letter we usually neglect backscattering, in order to properly compare with results that also neglected it, but we include backscattering for the CRESST underground run (see below).

We now compute the energy distribution of DM particles after $n$ scatterings. For spin-independent scattering, as long as the DM kinetic energy is low enough that the form factor is negligible, DM-nucleus scattering is isotropic in the c.m.\ frame, leading to a flat energy loss distribution:

\begin{equation}
    P(\Delta E) = \frac{1}{E_{max}}\theta(E_{max} - \Delta E)\,.
\end{equation}

A realistic overburden consists of multiple nuclei. In this case, the energy loss distribution after one collision is a sum over different nuclei, weighted by the relative probability of scattering with nucleus of type $A$:

\begin{equation}
    P(\Delta E) = \frac{\sum_A \frac{n_A \sigma_{\chi A}}{E_{max, A}} \theta(E_{max, A} - \Delta E)}{\sum_A n_A \sigma_{\chi A}}\,.
\end{equation}

Given an energy spectrum $\frac{dN}{dE}_{n-1}(E)$ before scattering, the spectrum after scattering, with the DM losing energy $\Delta E$ with probability $P(\Delta E)$, is found through a change of variables and integration over the probability distribution:

\begin{equation}
    \frac{dN}{dE}_n(E) =\int \left(\frac{E}{E-\Delta E}\right)\frac{dN}{dE}_{n-1}\left(\frac{E^2}{E-\Delta E}\right) \,P(\Delta E)\,d\Delta E\,.
\end{equation}

The energy spectrum of DM that reaches detector depth is then a weighted sum over the spectra after all scatterings:

\begin{equation}
    \frac{dN}{dE}_{total}(z,E) = \sum_{n=0}^{\infty}\int_z^{\infty}P_n(z')dz'\frac{dN}{dE}_n(E)\,.
\end{equation}
In practice, this sum can be truncated when additional scatterings produce a negligible flux above threshold. In this Letter, we truncate the sum at 300 scatterings, unless the cross section is small enough that over 90\% of the incoming DM flux either reaches detector depth or is scattered out of the atmosphere in fewer scatterings, in which case it is truncated sooner. With this energy spectrum, one can compute nuclear recoil rates in a detector, and determine whether that detector is sensitive to the specified DM mass and cross section.

\section{Comparison with Monte Carlo Results}

\begin{figure}
    \centering
    \includegraphics[width=\columnwidth]{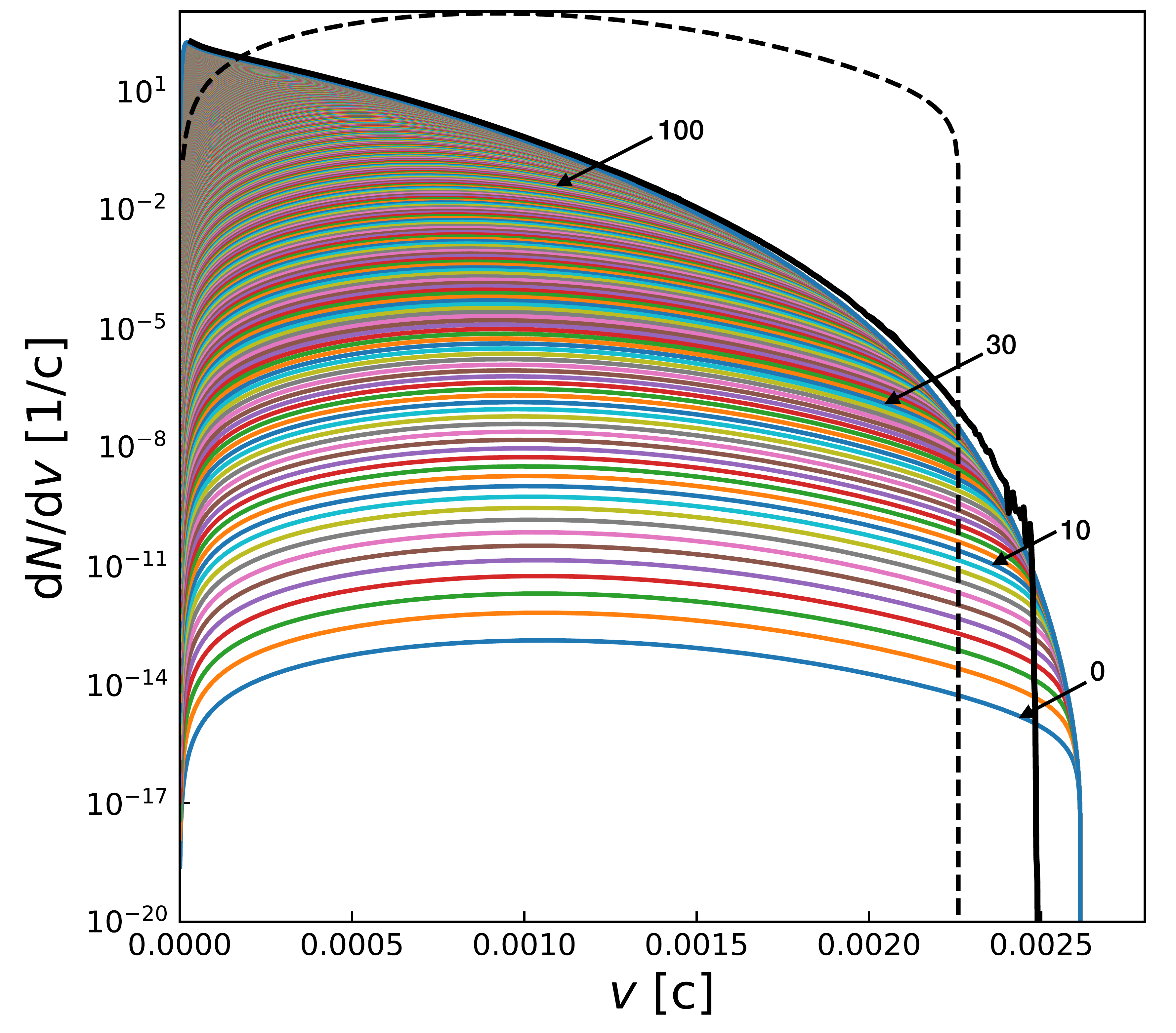}
    \caption{Velocity distribution of DM at a depth of 1400 m, normalized to the fraction of the incoming flux that reaches the detector, for $M_{\chi} = 100$ MeV and $\sigma_{\chi n} = 5\times10^{-30}$ cm$^2$. The bold black curve is the Monte Carlo result, dashed black is the straight-line approximation. The colored curves are the cumulative distributions of particles that scatter 0 to 1000 times, computed using the method of this Letter (number of scatterings is labeled for 4 example curves).}
    \label{fig:veldist}
\end{figure}

Figure~\ref{fig:veldist} shows the velocity distribution at a depth of 1400 meters for $M_{\chi} = $ 100 MeV and $\sigma_{\chi n} = 5\times10^{-30}$ cm$^2$. The multicolored curves show our computation of the cumulative velocity distributions of particles that scatter zero to 1000 times. That is, the lowest curve is the flux of particles that reach 1400 meters without scattering, the next is the flux of particles that scatter up to one time, etc. The solid black curve is the result of the DaMaSCUS-CRUST Monte Carlo code~\cite{Emken:2018run,DaMaSCUS-CRUST}. The high-velocity cutoff is not physical, but results from the simulation struggling to sample the extremely small flux at high velocity. For comparison, we also show the straight-line approximation used by, e.g., Refs.~\cite{Kavanagh:2017cru,EDELWEISS:2019vjv,McKeen:2022poo}. In this case, the high-velocity cutoff is a characteristic feature, as the method assumes that all particles of a given initial velocity lose the same amount of energy. The method of this Letter produces a much better fit than does the straight-line approximation.

\begin{figure}
    \centering
    \includegraphics[width=\columnwidth]{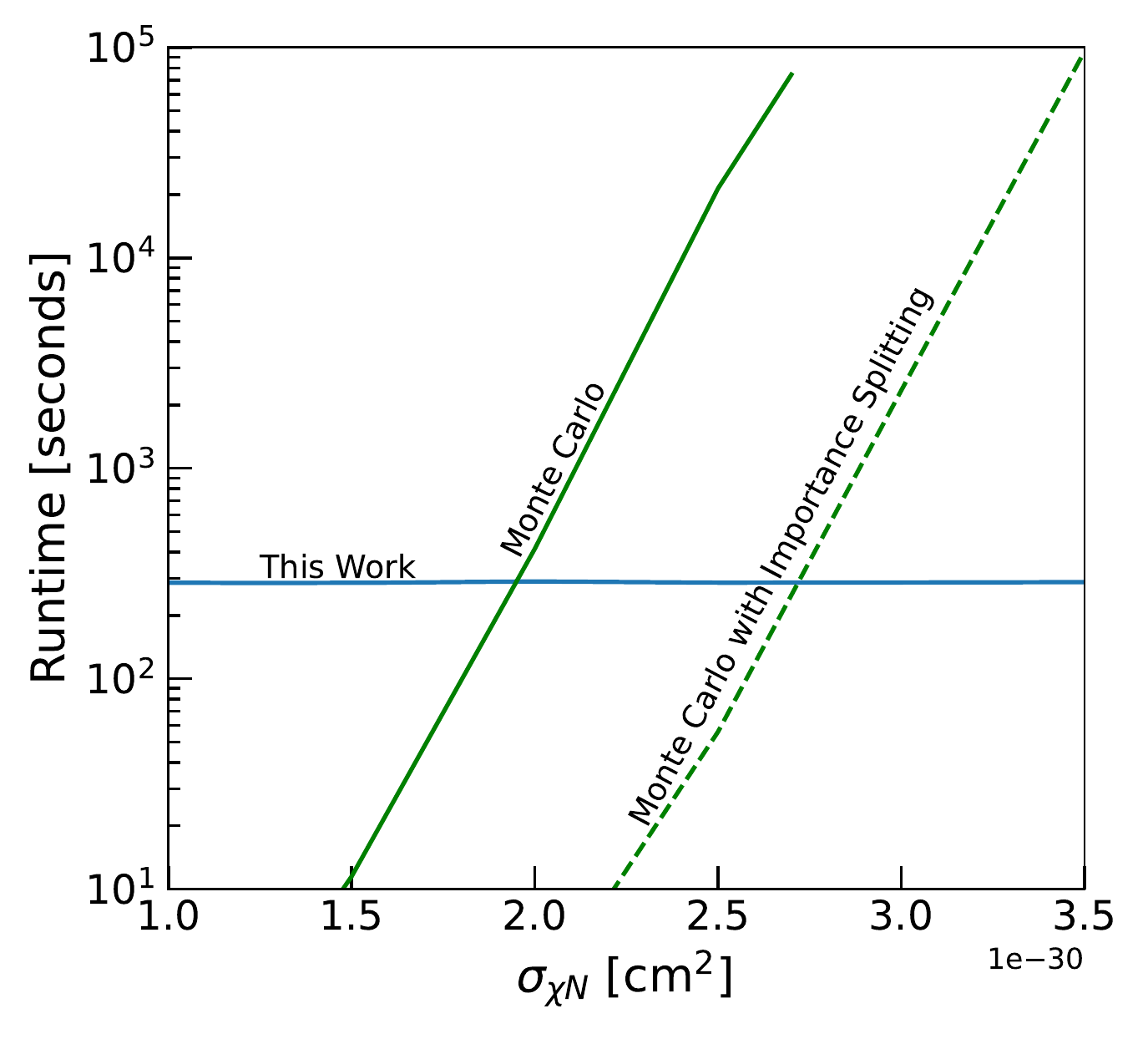}
    \caption{Comparison of the runtime for our method (blue) and Monte Carlo method (green), on a single core, for $M_{\chi} =$ 200 MeV and detector depth of 1400 m. The solid blue curve tracks DM through 100 scatterings. The solid green is the timing to obtain 100 particles able to trigger the CRESST-III detector with DaMaSCUS-CRUST, while the dashed green is the same using the importance splitting algorithm.}
    \label{fig:timing}
\end{figure}

Figure~\ref{fig:timing} compares the runtime of our method with that of the DaMaSCUS-CRUST Monte Carlo code~\cite{Emken:2018run,DaMaSCUS-CRUST}, for cross sections near the ceiling for the CRESST underground run (see Sec.~\ref{constraints}), and depth of 1400 meters. The blue line is the runtime needed for our method to model particles through 100 scatterings. Increasing this number to 300 or 1000 scatterings has little effect on the high-velocity end of the spectrum, because after so many scatterings most particles would have lost most of their energy. The green curves are the time needed to obtain a sample of 100 particles capable of triggering CRESST-III using DaMaSCUS-CRUST. The solid curve uses the default settings, while the dashed curve uses importance splitting (see Ref.~\cite{Emken:2019tni} for details). Although the Monte Carlo method is faster than ours at small cross sections, its runtime increases exponentially with $\sigma_{\chi n}$, while the runtime for ours is flat.

\section{Loosening Constraints on Sub-Dominant Dark Matter}\label{constraints}

Ceilings for numerous direct detection analyses have been computed using Monte Carlo and analytic methods~\cite{Davis:2017noy,Kavanagh:2017cru,Mahdawi:2017utm,Emken:2018run,Hooper:2018bfw,DEAPCollaboration:2021raj,EDELWEISS:2022ktt,SuperCDMS:2022kgp}. However, these analyses all assumed that the particle they consider makes up all of the DM ($f_{DM} = 1$). Recent papers on sub-dominant DM have not computed detector ceilings numerically, instead using approximate methods or taking ceilings from existing references. Focusing on the CRESST surface run and the CRESST-III underground analysis~\cite{CRESST:2017ues,CRESST:2019jnq}, we show that the ceiling of such an exclusion region can change significantly for $f_{DM} \ll 1$. As a baseline, we compare our results to the exclusion regions shown in Ref.~\cite{McKeen:2022poo}, for $f_{DM} = {1, 10^{-3}, \text{and}\, 10^{-6}}$.

\subsection{CRESST Surface Run}\label{cresstsurface}

\begin{figure}
    \centering
    \includegraphics[width=\columnwidth]{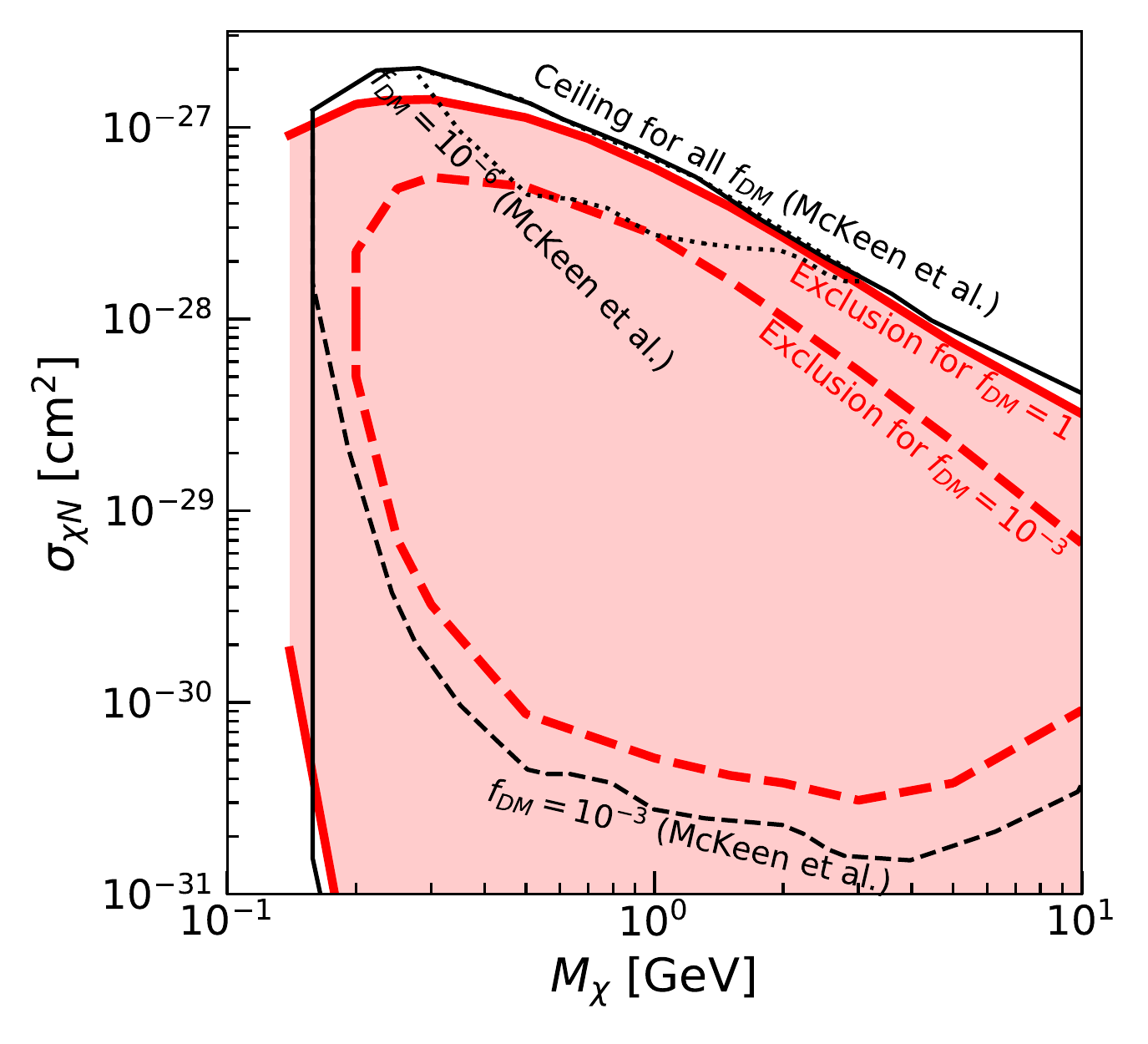}
    \caption{Solid black: exclusion region for the CRESST surface run for $f_{DM} = 1$, computed via Monte Carlo in Ref.~\cite{Emken:2018run} and used in Ref.~\cite{McKeen:2022poo}. Dashed and dotted black with a solid black ceiling are the exclusion regions for $f_{DM} = 10^{-3}$ and $10^{-6}$, respectively, reported in Ref.~\cite{McKeen:2022poo}. Solid and dashed red are our computed regions for $f_{DM} = 1$ and $10^{-3}$, respectively; we find that the CRESST surface run is not sensitive to $f_{DM} = 10^{-6}$. See Subsection~\ref{cresstsurface} for a discussion of why $f_{DM}$ affects the ceiling.}
    \label{fig:subdm}
\end{figure}

Figure~\ref{fig:subdm} shows our computation of the CRESST surface run exclusion region. We neglect backscattering from below detector depth, as was done in the Monte Carlo approach of Ref.~\cite{Emken:2018run}. The ceiling depends on the value of $f_{DM}$ because, at large cross sections, the probability of reaching the detector with high velocity becomes quite small (see Fig.~\ref{fig:veldist}). If the total DM flux is made smaller, then the probability of having high velocity must be made larger for the DM to be detectable, so the ceiling must be lower. The difference between the bottoms of our exclusion regions and those of Ref.~\cite{McKeen:2022poo} is that our treatment only considers DM arriving from above the horizon. The CRESST analysis (and by extension Ref.~\cite{McKeen:2022poo}) most likely assumed that the Earth was transparent to DM, which is reasonable near the bottom of their $f_{DM} = 1$ exclusion region. But for the parameter space shown, DM would have to scatter O(1000) times to cross the whole Earth, so it is reasonable to assume that $\sim$1/2 of the DM flux is blocked. In contrast with Ref.~\cite{McKeen:2022poo}, we find that the CRESST surface run cannot exclude parameter space for $f_{DM} = 10^{-6}$, because the flux is not sufficiently large to overcome attenuation.

There is a slight discrepancy between our ceiling for $f_{DM} = 1$ and the original Monte Carlo result of Ref.~\cite{Emken:2018run}. However, improvements to the DaMaSCUS-CRUST code have since lowered the Monte Carlo ceiling by a few tens of percent at low masses, bringing these lines into good agreement (\cite{privatecomm}, see recent updates to Ref.~\cite{DaMaSCUS-CRUST}). 

\subsection{CRESST Underground Limits}

Ref.~\cite{McKeen:2022poo} also computed the ceiling for the CRESST-III underground run~\cite{CRESST:2019jnq}, for $f_{DM} = 1$, using a straight-line approximation. Such approximations have been shown to be accurate~\cite{Emken:2018run} or even conservative ~\cite{Mahdawi:2017utm} when compared to numerical simulations, although Ref.~\cite{Emken:2019tni} comes to the opposite conclusion. Here we compute the ceiling for this detector using the method described in Sec.~\ref{code description}, including backscattering, for different values of $f_{DM}$.

\begin{figure}
    \centering
    \includegraphics[width=\columnwidth]{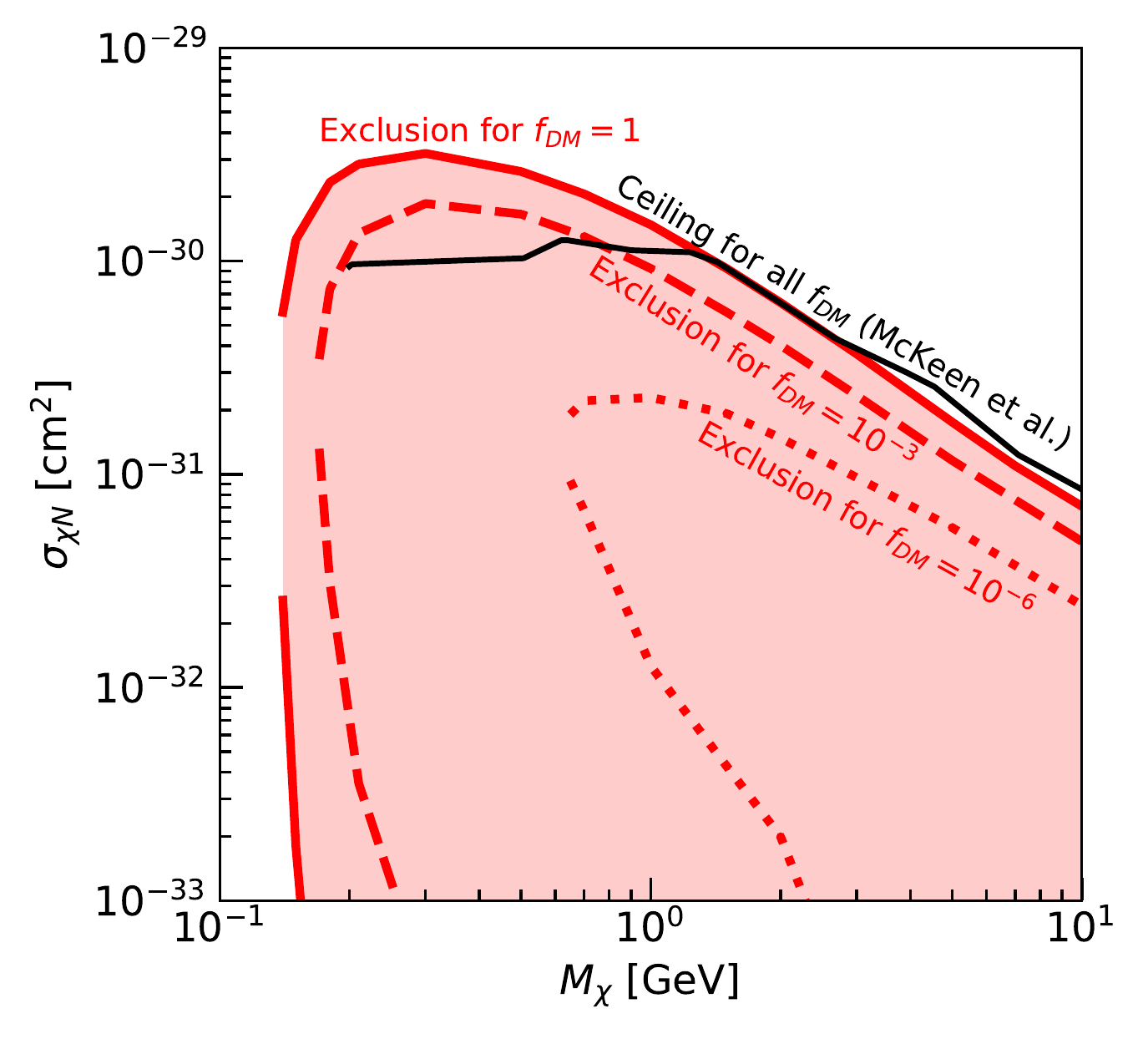}
    \caption{Solid red, shaded: parameter space excluded by the CRESST underground run for $f_{DM} = 1$. Dashed and dotted red: corresponding exclusion regions for $f_{DM} = 10^{-3}$ and $f_{DM} = 10^{-6}$, respectively. Black is the ballistic approximation result used in Ref.~\cite{McKeen:2022poo} as the ceiling for all $f_{DM}$.}
    \label{fig:cresstunderground}
\end{figure}

Figure~\ref{fig:cresstunderground} shows our computation of the CRESST-III exclusion region for $f_{DM} = 1$, $10^{-3}$, and $10^{-6}$, compared to the straight-line approximation for $f_{DM} = 1$. For $f_{DM} = 1$, we see that while the results agree at large mass, the straight-line result underestimates the ceiling by a factor of a few below about 1 GeV. We also see that for $f_{DM} = 10^{-6}$, the straight-line result overestimates the ceiling by a factor of several. This means that, for highly subdominant DM, underground detectors constrain less parameter space than the straight-line approximation would suggest. Note that at the ceiling for $f_{DM} = 1$, and $M_{\chi} = 200$ MeV, this Letter's method is faster than the Monte Carlo method with importance splitting by about an order of magnitude (see Fig.~\ref{fig:timing}).

In this Section, we have used the exclusion regions from Ref.~\cite{McKeen:2022poo} as a baseline for comparison, but our goal is not to criticize their approach. We only use them as an example of treatments that are common in the literature. In fact, our results show that the limits from Ref.~\cite{McKeen:2022poo} exceed the CRESST limits more than previously appreciated.

\section{Conclusions}

When searching for dark matter, it is crucial to understand what parameter space has already been excluded, in order to effectively design new searches and interpret existing results. However, treating attenuation with a detailed Monte Carlo approach can be computationally intensive, leading to the use of approximate methods or simplified rescaling of existing limits. In this Letter, we present a novel method for modeling dark matter attenuation, which is more appropriate than the straight-line approximation for light dark matter, and faster than Monte Carlo methods for large cross sections. We reanalyze constraints on subdominant dark matter from multiple CRESST detectors, showing that they exclude less parameter space than previously thought.

As new technologies and analysis methods allow ever-lower energy thresholds, detectors will become sensitive to lighter dark matter, and to particles that have lost more energy in propagation. This will necessitate propagation methods that are valid for light dark matter, and feasible in the limit of many scatterings, two limits which our method is tailor made to handle. In principle, our approach can also be applied to DM interactions in neutron stars and white dwarfs, where the high density leads to many scatterings even for small cross sections. It could also be applied to other physical processes, such as neutrons propagating through dense media.  

\begin{acknowledgments}
I am extremely grateful to Timon Emken for detailed discussions about the DaMaSCUS Monte Carlo code, and to Ivan Esteban for suggestions for optimizing the integration performed in this work. I also thank Joseph Bramante, Marianne Moore, David Morrissey, and Aaron Vincent for helpful comments and discussions, and the anonymous referee for comments which made the text more clear. Computations were performed on equipment funded by the Canada Foundation for Innovation and the Ontario Government and operated by the Queen’s Centre for Advanced Computing. C.V.C. was supported by the Arthur B. McDonald Canadian Astroparticle Physics Research Institute. Research at Perimeter Institute is supported by the Government of Canada through the Department of Innovation, Science, and Economic Development, and by the Province of Ontario.
\end{acknowledgments}

\bibliography{main}


\end{document}